\documentstyle[10pt,epsf,epsfig,dp_delphititle]{dp_delphi}
\makeindex
\def\DpPaperGroup{EP}
\def\DpPaperRef{2000-020}
\def\DpDate{31 January 2000}
\def\DpAuthors{DELPHI Collaboration}
\def\DpSubmit{(Phys. Lett. B478(2000)65)}
\def\DpTitle{{\bf Search for Heavy Stable and Long-Lived Particles in
\boldmath e$^{+}$e$^{-}$ Collisions at $\sqrt{s}$=189 GeV }}
\def\DpComment{ }
\def\DpEMail{ }

\begin{document}
\makeatletter
\newcount\@tempcntc
\def\@citex[#1]#2{\if@filesw\immediate\write\@auxout{\string\citation{#2}}\fi
  \@tempcnta\z@\@tempcntb\m@ne\def\@citea{}\@cite{\@for\@citeb:=#2\do
    {\@ifundefined
       {b@\@citeb}{\@citeo\@tempcntb\m@ne\@citea\def\@citea{,}{\bf ?}\@warning
       {Citation `\@citeb' on page \thepage \space undefined}}%
    {\setbox\z@\hbox{\global\@tempcntc0\csname b@\@citeb\endcsname\relax}%
     \ifnum\@tempcntc=\z@ \@citeo\@tempcntb\m@ne
       \@citea\def\@citea{,}\hbox{\csname b@\@citeb\endcsname}%
     \else
      \advance\@tempcntb\@ne
      \ifnum\@tempcntb=\@tempcntc
      \else\advance\@tempcntb\m@ne\@citeo
      \@tempcnta\@tempcntc\@tempcntb\@tempcntc\fi\fi}}\@citeo}{#1}}
\def\@citeo{\ifnum\@tempcnta>\@tempcntb\else\@citea\def\@citea{,}%
  \ifnum\@tempcnta=\@tempcntb\the\@tempcnta\else
   {\advance\@tempcnta\@ne\ifnum\@tempcnta=\@tempcntb \else \def\@citea{--}\fi
    \advance\@tempcnta\m@ne\the\@tempcnta\@citea\the\@tempcntb}\fi\fi}
 
\makeatother
\begin{titlepage}
\pagenumbering{roman}
\CERNpreprint{\DpPaperGroup}{\DpPaperRef} 
\date{{\small\DpDate}} 
\title{\DpTitle} 
\address{\DpAuthors} 
\begin{shortabs} 
\noindent
%
\noindent

A search for stable and long-lived heavy charged particles 
was performed  using the data taken by the DELPHI experiment 
at an energy of 189 GeV.
The Cherenkov light detected in the Ring Imaging Cherenkov Detector
and the ionisation loss measured in the Time Projection Chamber were used to
identify heavy particles passing through the detector. No evidence for the
production of such particles has been found, therefore 
exclusion limits at 95\% confidence level were derived 
on the masses of left and right handed smuons and staus.
The results were combined with previous DELPHI searches in this channel.
Including previous DELPHI results, masses of 
left (right) handed stable smuons and staus can be excluded
between 2 GeV/c$^2$ and 88 (87.5)~GeV/c$^2$ at 95\% CL.

\end{shortabs}
\vfill
\begin{center}
\DpSubmit \ \\ 
\DpComment \ \\
\DpEMail \ \\
\end{center}
\vfill
\clearpage
\headsep 10.0pt
\addtolength{\textheight}{10mm}
\addtolength{\footskip}{-5mm}
\begingroup
%
\newcommand{\DpName}[2]{\hbox{#1$^{\ref{#2}}$},\hfill}
\newcommand{\DpNameTwo}[3]{\hbox{#1$^{\ref{#2},\ref{#3}}$},\hfill}
\newcommand{\DpNameThree}[4]{\hbox{#1$^{\ref{#2},\ref{#3},\ref{#4}}$},\hfill}
\newskip\Bigfill \Bigfill = 0pt plus 1000fill
\newcommand{\DpNameLast}[2]{\hbox{#1$^{\ref{#2}}$}\hspace{\Bigfill}}
%
\footnotesize
\noindent
\DpName{P.Abreu}{LIP}
\DpName{W.Adam}{VIENNA}
\DpName{T.Adye}{RAL}
\DpName{P.Adzic}{DEMOKRITOS}
\DpName{Z.Albrecht}{KARLSRUHE}
\DpName{T.Alderweireld}{AIM}
\DpName{G.D.Alekseev}{JINR}
\DpName{R.Alemany}{VALENCIA}
\DpName{T.Allmendinger}{KARLSRUHE}
\DpName{P.P.Allport}{LIVERPOOL}
% ...
\normalsize
\endgroup
\titlefoot{Department of Physics and Astronomy, Iowa State
     University, Ames IA 50011-3160, USA
    \label{AMES}}
\titlefoot{Physics Department, Univ. Instelling Antwerpen,
     Universiteitsplein 1, B-2610 Antwerpen, Belgium \\
     \indent~~and IIHE, ULB-VUB,
     Pleinlaan 2, B-1050 Brussels, Belgium \\
     \indent~~and Facult\'e des Sciences,
     Univ. de l'Etat Mons, Av. Maistriau 19, B-7000 Mons, Belgium
    \label{AIM}}
\titlefoot{Physics Laboratory, University of Athens, Solonos Str.
     104, GR-10680 Athens, Greece
    \label{ATHENS}}
\titlefoot{Department of Physics, University of Bergen,
     All\'egaten 55, NO-5007 Bergen, Norway
    \label{BERGEN}}
\titlefoot{Dipartimento di Fisica, Universit\`a di Bologna and INFN,
     Via Irnerio 46, IT-40126 Bologna, Italy
    \label{BOLOGNA}}
\titlefoot{Centro Brasileiro de Pesquisas F\'{\i}sicas, rua Xavier Sigaud 150,
     BR-22290 Rio de Janeiro, Brazil \\
     \indent~~and Depto. de F\'{\i}sica, Pont. Univ. Cat\'olica,
     C.P. 38071 BR-22453 Rio de Janeiro, Brazil \\
     \indent~~and Inst. de F\'{\i}sica, Univ. Estadual do Rio de Janeiro,
     rua S\~{a}o Francisco Xavier 524, Rio de Janeiro, Brazil
    \label{BRASIL}}
\titlefoot{Comenius University, Faculty of Mathematics and Physics,
     Mlynska Dolina, SK-84215 Bratislava, Slovakia
    \label{BRATISLAVA}}
\titlefoot{Coll\`ege de France, Lab. de Physique Corpusculaire, IN2P3-CNRS,
     FR-75231 Paris Cedex 05, France
    \label{CDF}}
\titlefoot{CERN, CH-1211 Geneva 23, Switzerland
    \label{CERN}}
\titlefoot{Institut de Recherches Subatomiques, IN2P3 - CNRS/ULP - BP20,
     FR-67037 Strasbourg Cedex, France
    \label{CRN}}
\titlefoot{Now at DESY-Zeuthen, Platanenallee 6, D-15735 Zeuthen, Germany
    \label{DESY}}
\titlefoot{Institute of Nuclear Physics, N.C.S.R. Demokritos,
     P.O. Box 60228, GR-15310 Athens, Greece
    \label{DEMOKRITOS}}
\titlefoot{FZU, Inst. of Phys. of the C.A.S. High Energy Physics Division,
     Na Slovance 2, CZ-180 40, Praha 8, Czech Republic
    \label{FZU}}
\titlefoot{Dipartimento di Fisica, Universit\`a di Genova and INFN,
     Via Dodecaneso 33, IT-16146 Genova, Italy
    \label{GENOVA}}
\titlefoot{Institut des Sciences Nucl\'eaires, IN2P3-CNRS, Universit\'e
     de Grenoble 1, FR-38026 Grenoble Cedex, France
    \label{GRENOBLE}}
\titlefoot{Helsinki Institute of Physics, HIP,
     P.O. Box 9, FI-00014 Helsinki, Finland
    \label{HELSINKI}}
\titlefoot{Joint Institute for Nuclear Research, Dubna, Head Post
     Office, P.O. Box 79, RU-101 000 Moscow, Russian Federation
    \label{JINR}}
\titlefoot{Institut f\"ur Experimentelle Kernphysik,
     Universit\"at Karlsruhe, Postfach 6980, DE-76128 Karlsruhe,
     Germany
    \label{KARLSRUHE}}
\titlefoot{Institute of Nuclear Physics and University of Mining and Metalurgy,
     Ul. Kawiory 26a, PL-30055 Krakow, Poland
    \label{KRAKOW}}
\titlefoot{Universit\'e de Paris-Sud, Lab. de l'Acc\'el\'erateur
     Lin\'eaire, IN2P3-CNRS, B\^{a}t. 200, FR-91405 Orsay Cedex, France
    \label{LAL}}
\titlefoot{School of Physics and Chemistry, University of Lancaster,
     Lancaster LA1 4YB, UK
    \label{LANCASTER}}
\titlefoot{LIP, IST, FCUL - Av. Elias Garcia, 14-$1^{o}$,
     PT-1000 Lisboa Codex, Portugal
    \label{LIP}}
\titlefoot{Department of Physics, University of Liverpool, P.O.
     Box 147, Liverpool L69 3BX, UK
    \label{LIVERPOOL}}
\titlefoot{LPNHE, IN2P3-CNRS, Univ.~Paris VI et VII, Tour 33 (RdC),
     4 place Jussieu, FR-75252 Paris Cedex 05, France
    \label{LPNHE}}
\titlefoot{Department of Physics, University of Lund,
     S\"olvegatan 14, SE-223 63 Lund, Sweden
    \label{LUND}}
\titlefoot{Universit\'e Claude Bernard de Lyon, IPNL, IN2P3-CNRS,
     FR-69622 Villeurbanne Cedex, France
    \label{LYON}}
\titlefoot{Univ. d'Aix - Marseille II - CPP, IN2P3-CNRS,
     FR-13288 Marseille Cedex 09, France
    \label{MARSEILLE}}
\titlefoot{Dipartimento di Fisica, Universit\`a di Milano and INFN-MILANO,
     Via Celoria 16, IT-20133 Milan, Italy
    \label{MILANO}}
\titlefoot{Dipartimento di Fisica, Univ. di Milano-Bicocca and
     INFN-MILANO, Piazza delle Scienze 2, IT-20126 Milan, Italy
    \label{MILANO2}}
\titlefoot{Niels Bohr Institute, Blegdamsvej 17,
     DK-2100 Copenhagen {\O}, Denmark
    \label{NBI}}
\titlefoot{IPNP of MFF, Charles Univ., Areal MFF,
     V Holesovickach 2, CZ-180 00, Praha 8, Czech Republic
    \label{NC}}
\titlefoot{NIKHEF, Postbus 41882, NL-1009 DB
     Amsterdam, The Netherlands
    \label{NIKHEF}}
\titlefoot{National Technical University, Physics Department,
     Zografou Campus, GR-15773 Athens, Greece
    \label{NTU-ATHENS}}
\titlefoot{Physics Department, University of Oslo, Blindern,
     NO-1000 Oslo 3, Norway
    \label{OSLO}}
\titlefoot{Dpto. Fisica, Univ. Oviedo, Avda. Calvo Sotelo
     s/n, ES-33007 Oviedo, Spain
    \label{OVIEDO}}
\titlefoot{Department of Physics, University of Oxford,
     Keble Road, Oxford OX1 3RH, UK
    \label{OXFORD}}
\titlefoot{Dipartimento di Fisica, Universit\`a di Padova and
     INFN, Via Marzolo 8, IT-35131 Padua, Italy
    \label{PADOVA}}
\titlefoot{Rutherford Appleton Laboratory, Chilton, Didcot
     OX11 OQX, UK
    \label{RAL}}
\titlefoot{Dipartimento di Fisica, Universit\`a di Roma II and
     INFN, Tor Vergata, IT-00173 Rome, Italy
    \label{ROMA2}}
\titlefoot{Dipartimento di Fisica, Universit\`a di Roma III and
     INFN, Via della Vasca Navale 84, IT-00146 Rome, Italy
    \label{ROMA3}}
\titlefoot{DAPNIA/Service de Physique des Particules,
     CEA-Saclay, FR-91191 Gif-sur-Yvette Cedex, France
    \label{SACLAY}}
\titlefoot{Instituto de Fisica de Cantabria (CSIC-UC), Avda.
     los Castros s/n, ES-39006 Santander, Spain
    \label{SANTANDER}}
\titlefoot{Dipartimento di Fisica, Universit\`a degli Studi di Roma
     La Sapienza, Piazzale Aldo Moro 2, IT-00185 Rome, Italy
    \label{SAPIENZA}}
\titlefoot{Inst. for High Energy Physics, Serpukov
     P.O. Box 35, Protvino, (Moscow Region), Russian Federation
    \label{SERPUKHOV}}
\titlefoot{J. Stefan Institute, Jamova 39, SI-1000 Ljubljana, Slovenia
     and Laboratory for Astroparticle Physics,\\
     \indent~~Nova Gorica Polytechnic, Kostanjeviska 16a, SI-5000 Nova Gorica, Slovenia, \\
     \indent~~and Department of Physics, University of Ljubljana,
     SI-1000 Ljubljana, Slovenia
    \label{SLOVENIJA}}
\titlefoot{Fysikum, Stockholm University,
     Box 6730, SE-113 85 Stockholm, Sweden
    \label{STOCKHOLM}}
\titlefoot{Dipartimento di Fisica Sperimentale, Universit\`a di
     Torino and INFN, Via P. Giuria 1, IT-10125 Turin, Italy
    \label{TORINO}}
\titlefoot{Dipartimento di Fisica, Universit\`a di Trieste and
     INFN, Via A. Valerio 2, IT-34127 Trieste, Italy \\
     \indent~~and Istituto di Fisica, Universit\`a di Udine,
     IT-33100 Udine, Italy
    \label{TU}}
\titlefoot{Univ. Federal do Rio de Janeiro, C.P. 68528
     Cidade Univ., Ilha do Fund\~ao
     BR-21945-970 Rio de Janeiro, Brazil
    \label{UFRJ}}
\titlefoot{Department of Radiation Sciences, University of
     Uppsala, P.O. Box 535, SE-751 21 Uppsala, Sweden
    \label{UPPSALA}}
\titlefoot{IFIC, Valencia-CSIC, and D.F.A.M.N., U. de Valencia,
     Avda. Dr. Moliner 50, ES-46100 Burjassot (Valencia), Spain
    \label{VALENCIA}}
\titlefoot{Institut f\"ur Hochenergiephysik, \"Osterr. Akad.
     d. Wissensch., Nikolsdorfergasse 18, AT-1050 Vienna, Austria
    \label{VIENNA}}
\titlefoot{Inst. Nuclear Studies and University of Warsaw, Ul.
     Hoza 69, PL-00681 Warsaw, Poland
    \label{WARSZAWA}}
\titlefoot{Fachbereich Physik, University of Wuppertal, Postfach
     100 127, DE-42097 Wuppertal, Germany
    \label{WUPPERTAL}}
\addtolength{\textheight}{-10mm}
\addtolength{\footskip}{5mm}
\clearpage
\headsep 30.0pt
\end{titlepage}
%
\pagenumbering{arabic} 
\setcounter{footnote}{0} %
\large
%
\section{Introduction}

The search for heavy stable and long-lived charged particles in events with low
multiplicity was made within the data recorded by the DELPHI detector at LEP at
center-of-mass energies of about 189~GeV. The results presented in this paper extend previous
DELPHI results from \cite{ref:stable} in the leptonic topology to energies of 189 GeV.
In most Supersymmetric models (SUSY) the supersymmetric partners of standard
particles are unstable and have short lifetimes, except the lightest  supersymmetric
particle (LSP) which is commonly believed to be neutral and stable.
Therefore, in most of the searches it is assumed that the 
supersymmetric particles decay instantaneously. 
In some scenarious, however, it is possible that SUSY-particles acquire a long lifetime.
They become long-lived or stable, and thus escape from detection by the standard searches.
In the Minimal Supersymmetric Standard Model (MSSM)
with a very small amount of R-parity violation 
the LSP can be a charged slepton or squark and decay with a long lifetime 
into Standard Model particles \cite{ref:rpar}.

In gauge mediated supersymmetric models \cite{ref:giu} the gravitino
is the LSP and the next to lightest supersymmetric particle 
(NLSP) can obtain a long lifetime in a very natural way 
for large values of the SUSY-breaking scale \cite{ref:slepton}.
This is possible for sleptons, for example when the stau is the NLSP.
The pair production of such long-lived or stable particles 
yields a characteristic signature,  with typically two back-to-back
charged heavy objects in the detector.

In this analysis it was assumed that the long-lived particles 
decay outside the tracking volume of the detector,
which extends to a typical radius of 1.5 m. Furthermore, it was assumed
that they do not interact more strongly than ordinary matter particles.

 Heavy stable particles were searched for by looking for high momentum
charged particles with either anomalous ionisation loss (dE/dx) 
measured in the Time Projection Chamber (TPC), or the absence of 
Cherenkov light in the gas and liquid radiators of 
the Barrel Ring Imaging CHerenkov (RICH).
The combination of the TPC and RICH detectors and kinematical
cuts provided efficient detection of new heavy particles with 
a small background.

The data analysed corresponded to 155.3 pb$^{-1}$ at an energy of
about 189 GeV.
The negative search result was used to set 
limits at 95\% CL on the production cross-section of
stable sleptons. They were combined with previous DELPHI results
from \cite{ref:stable}, and mass limits on stable or long-lived smuons
and staus were derived.

\section{Event selection}

 A description of the DELPHI apparatus and its performance can be 
found in ref.~\cite{ref:delp}, with more details on the Barrel RICH in 
ref.~\cite{ref:brich}.
 Charged particles were selected if their impact parameter was less than 5 cm in the 
azimuthal plane and less than 10 cm in the longitudinal direction,
 and their polar angles lay between 20$^\circ$ and 160$^\circ$.
The relative errors on the measured momentum\nolinebreak\footnote{Within this paper, with "momentum" the apparent momentum is meant, defined as
the ratio of the momentum and the charge $|q|$.} were required to be less than 100\%
and the track lengths larger than 30 cm.

Only events with two or three charged particles were considered. 
It was further required that at least one charged particle had a momentum 
above 5 GeV/c reconstructed by the TPC and was 
inside the acceptance of the Barrel RICH $|\cos\theta|<0.68$,
where $\theta$ is the polar angle.

 Cosmic muons were removed by putting tighter cuts on the impact parameter 
with respect to the average beam-spot position.
When the event had two charged particles with at least one 
identified muon in the muon chambers, the impact parameter 
in the azimuthal plane was required to be less than 0.15 cm, and less than 1.5 cm
in the longitudinal plane. 

Charged particles were selected, if they fullfilled a combination of the
following criteria:
\begin{enumerate}
 \item the Gas Veto:  no photons were observed in the Gas RICH
 \item the Liquid Veto:  four or less photons were observed in 
the Liquid RICH
 \item the TPC high ionisation loss: measured normalised energy loss 
was above 2 units i.e. twice the energy loss for a minimum ionising particle
 \item the TPC low ionisation loss: measured normalised energy loss 
was below 0.3 of that expected for protons
\end{enumerate}
The particle identification using the RICH is described in 
detail in ref. \cite{ref:ribmean}.

 For the Gas and Liquid Vetoes it was required that the RICH was 
fully operational, and that photons from other tracks or
ionisation hits were detected inside the drift tube crossed by the track.
Due to tracking problems electrons often passed a Gas or Liquid Veto.
Therefore it was required that particles that deposit more than 5 GeV 
in the electromagnetic calorimeter, had hits in the outer tracking detector.
Signals from at least 80 wires were required for the measurement 
of the normalised energy loss in the TPC.

 An event was selected if the momentum of a charged 
particle was above 15 GeV/c and the Gas Veto (1) was confirmed by a Liquid Veto (2) or 
a low ionisation loss (4) (in boolean notation (1)$\cdot$(2)+(1)$\cdot$(4)) or if 
the momentum of a charged particle 
was above 5 GeV/c and the Gas Veto was confirmed by a high ionisation loss 
((1)$\cdot$(3)). The event was also accepted if two 
charged particles were present with both momenta 
above 15 GeV/c,
and both had either a high ionisation loss or a Gas Veto, or both had a
low ionisation loss ((1+3)$\cdot$(1'+3')+(4)$\cdot$(4')).
This additional search window improves the efficiency
for large masses by about 10\%.

\section{Analysis results}

One event was selected in the data, in agreement with
the expected background of 1.02 $\pm$ 0.13  events.
The expected background was evaluated from the data by
estimating the probability of accepted particles passing
the individual cuts which build up the different search windows.
The largest background contribution was expected in the high energy loss search window,
0.54$\pm$ 0.1. A background of 0.34$\pm 0.07$ was expected in the combined Liquid and Gas
Veto search window, whereas the contribution from the other two windows was small.
The candidate event passed both the Gas and Liquid Vetoes. It had two high momentum
back-to-back charged particles with associated hits in the muon chambers. 
Figure 1 shows the measured normalised ionisation loss and the
measured Cherenkov angle in the liquid radiator for the data taken at 189 GeV,
after applying the Gas Veto.

The efficiency for selecting an event was evaluated  for right handed smuons
as a function of the mass in 10~GeV/c$^2$ steps between 10~GeV/c$^2$ and 90~GeV/c$^2$.
Additional points at 85,91,92 and 93~GeV/c$^2$ have been added in order to get a
better modelling of the efficiency close to the kinematical limit. At each point
1000 events were simulated with SUSYGEN \cite{ref:stavros}, and passed through
the detector simulation.
The efficiency curve for masses above 45~GeV/c$^2$ and a centre-of-mass energy
of 189 GeV is shown in Figure 2a. Stable or long-lived staus and selectrons are expected to yield the
same experimental signature, so the estimated efficiencies hold for these cases as well.

As no evidence for a signal was found, exclusion limits at 95\% CL were derived.
The new results were combined with previous
published results from \cite{ref:stable} with a likelihood ratio method
to obtain experimental limits on the masses of left and right
handed stable or long-lived smuons and staus.
For selectrons, no limits were given because in this model the production cross-section
can be highly suppressed due to an addition t-channel sneutrino exchange.
The candidate event has been
taken into account over the full mass range. 
Figure 2b shows the cross-section for left and right handed smuons in the MSSM 
at an energy of 189 GeV, calculated with SUSYGEN \cite{ref:stavros}
as a function of the particle mass. The 95\% CL cross-section
limit is also shown. 
The production of left (right) handed stable smuons and staus can be excluded
with masses up to 88 (87.5)~GeV/c$^2$. 

\section{Conclusions}

A search was made for stable and long-lived heavy charged particles 
in leptonic final states 
at energies of 189 GeV, using particles identified 
by the Cherenkov light in the RICH and the 
ionisation loss in the TPC.

One event was observed in the data, in agreement with the expectation of
1.02 $\pm$ 0.13 background events. The results were combined with previous results of
the  DELPHI Collaboration\cite{ref:stable}, which excluded long lived or stable
left (right) handed smuons and staus for masses between 2 and 81 (80)~GeV/c$^2$. 
Including the new data,
left (right) handed stable smuons and staus can be excluded
with masses between 2 GeV/c$^2$ and 88 (87.5)~GeV/c$^2$ at 95\% CL. 

\subsection*{Acknowledgements}
\vskip 3 mm
 We are greatly indebted to our technical 
collaborators, to the members of the CERN-SL Division for the excellent 
performance of the LEP collider, and to the funding agencies for their
support in building and operating the DELPHI detector.\\
We acknowledge in particular the support of \\
Austrian Federal Ministry of Science and Traffics, GZ 616.364/2-III/2a/98, \\
FNRS--FWO, Belgium,  \\
FINEP, CNPq, CAPES, FUJB and FAPERJ, Brazil, \\
Czech Ministry of Industry and Trade, GA CR 202/96/0450 and GA AVCR A1010521,\\
Danish Natural Research Council, \\
Commission of the European Communities (DG XII), \\
Direction des Sciences de la Mati$\grave{\mbox{\rm e}}$re, CEA, France, \\
Bundesministerium f$\ddot{\mbox{\rm u}}$r Bildung, Wissenschaft, Forschung 
und Technologie, Germany,\\
General Secretariat for Research and Technology, Greece, \\
National Science Foundation (NWO) and Foundation for Research on Matter (FOM),
The Netherlands, \\
Norwegian Research Council,  \\
State Committee for Scientific Research, Poland, 2P03B06015, 2P03B1116 and
SPUB/P03/178/98, \\
JNICT--Junta Nacional de Investiga\c{c}\~{a}o Cient\'{\i}fica 
e Tecnol$\acute{\mbox{\rm o}}$gica, Portugal, \\
Vedecka grantova agentura MS SR, Slovakia, Nr. 95/5195/134, \\
Ministry of Science and Technology of the Republic of Slovenia, \\
CICYT, Spain, AEN96--1661 and AEN96-1681,  \\
The Swedish Natural Science Research Council,      \\
Particle Physics and Astronomy Research Council, UK, \\
Department of Energy, USA, DE--FG02--94ER40817. \\


\begin{figure}[htb]
\begin{center}
\mbox{\epsfxsize17.0cm 
\epsffile{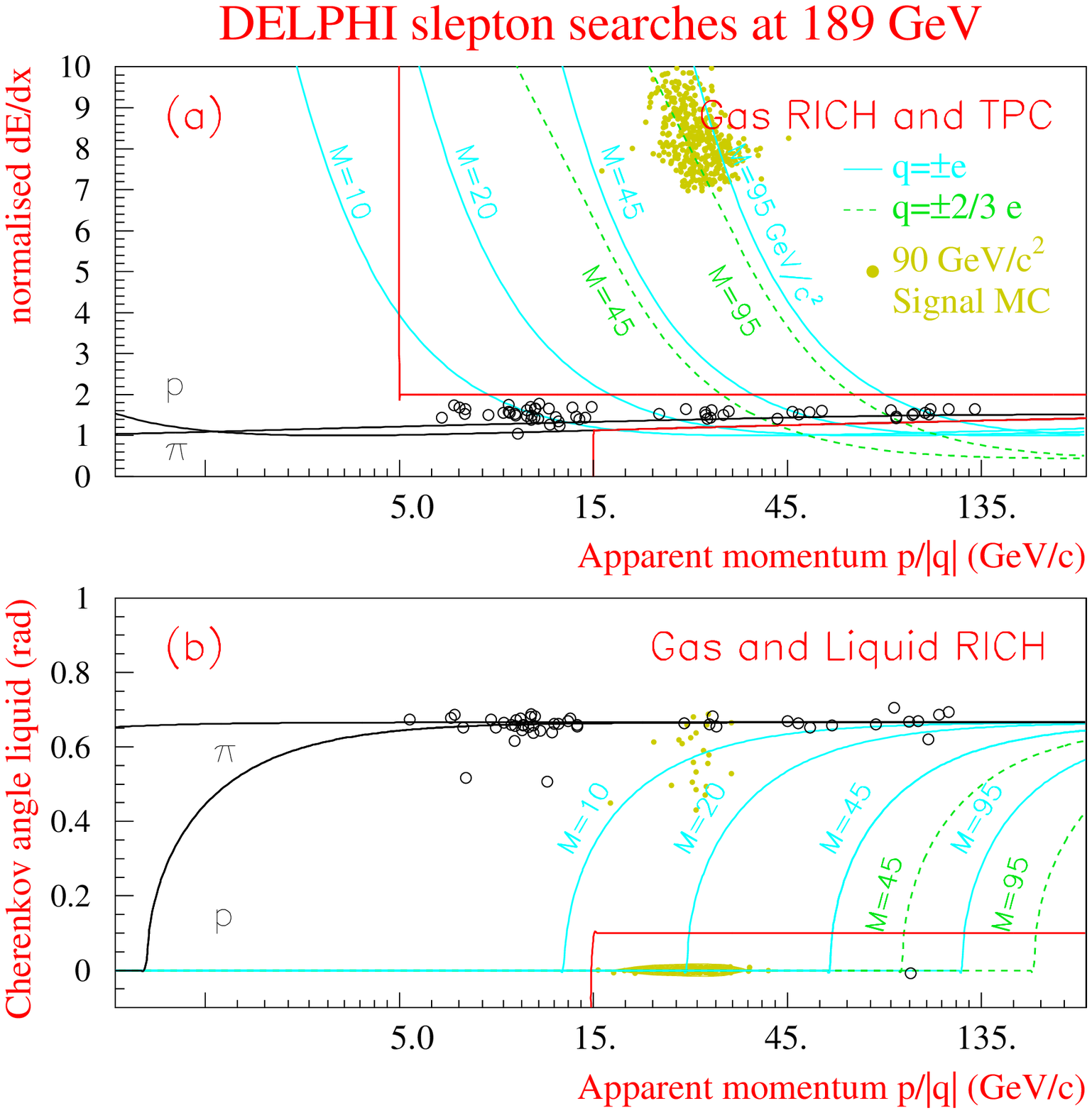} }
\end{center}
\vspace*{0.cm}
\caption{
(a) Normalised energy loss as a function of the apparent momentum 
p/$|q|$ after the Gas Veto for the 189 GeV data.
(b) Measured Cherenkov angle in the liquid radiator 
as a function of the apparent momentum after the Gas Veto:
if four photons or less were observed 
in the liquid radiator, the Cherenkov angle was set equal to zero.
The rectangular areas in (a) indicate selections (1)$\cdot$(2) and (1)$\cdot$(4),
and that in (b) shows selection (1)$\cdot$(3). The selection criteria are
explained in the text. Open circles are data. The small filled circles
indicate the expectation for a 90~GeV/c$^2$ signal with charge $\pm$e,
resulting in a large dE/dx (upper
plot) and no photons (except for a few accidental rings) in the liquid
Cherenkov counter (lower plot).}
\end{figure}

\begin{figure}[htb]
\begin{center}
\mbox{\epsfxsize15.0cm 
\epsffile{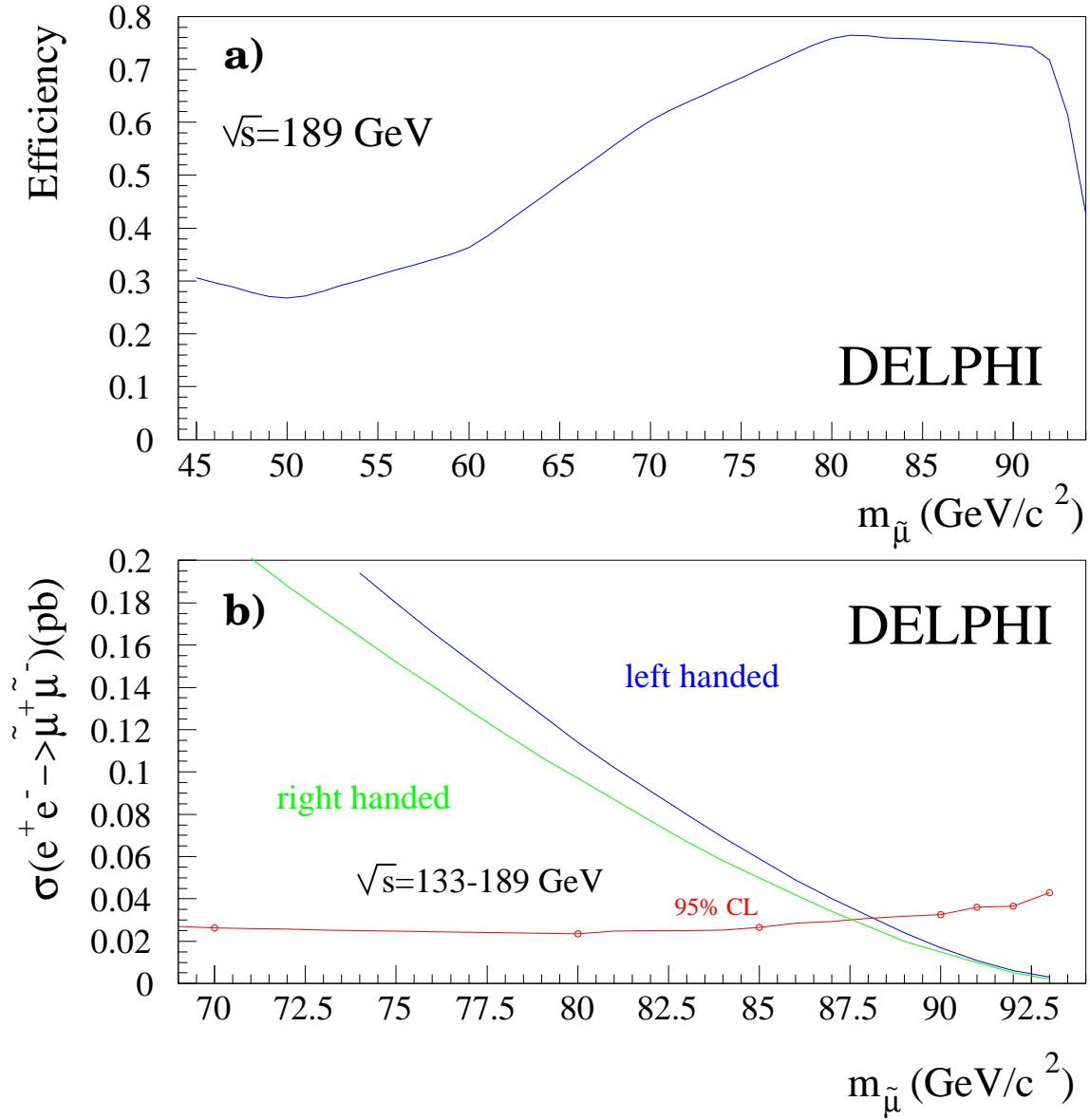} }
\end{center}
\vspace*{0.cm}
\caption{(a) Efficiency for detecting stable and long-lived smuons (staus)
as a function of the smuon mass at a centre-of-mass energy of 189 GeV.\newline
(b)Predicted production cross-section for left and right handed smuons (staus) as
a function of the smuon (stau) mass. The 95\% CL cross-section limit, derived from
DELPHI searches between 133 GeV and 189 GeV, is also shown.}
\end{figure}

\end{document}